\documentclass[reprint,amssymb,prl,longbibliography]{revtex4-2}
\usepackage[utf8]{inputenc}
\usepackage[T1]{fontenc}
\usepackage{graphicx}
\usepackage{amsmath}
\usepackage{amssymb}
\usepackage{mathrsfs}
\usepackage{braket}
\usepackage{geometry}
\usepackage{here}
\usepackage{lmodern}
\usepackage{array}
\usepackage{natbib}
\usepackage{url}
\usepackage{textcase}
\usepackage{bm}
\usepackage{natbib}
\usepackage{color}
\begin{document}
	\title{Robust Coherent Control of Bimolecular Collisions beyond the Ultracold Regime}

\author{Adrien Devolder$^{1}$, Paul Brumer$^{1}$, and Timur V. Tscherbul$^{2}$}

\affiliation{$^{1}$Chemical Physics Theory Group, Department of Chemistry, and Center for Quantum Information and Quantum Control, University of Toronto, Toronto, Ontario, M5S 3H6, Canada\\
	$^{2}$Department of Physics, University of Nevada, Reno, NV, 89557, USA}

\begin{abstract}
	Quantum coherent control of bimolecular collisions beyond the ultracold regime can face a major challenge due to the incoherent addition of different partial wave contributions to the total scattering cross section. These contributions become increasingly numerous as the collision energy increases, leading to a loss of overall control. Here, we overcome this limitation by leveraging the recently discovered Partial Wave Phase Locking (PWPL) effect, which synchronizes the oscillations of all partial wave contributions. By using rigorous quantum scattering calculations, we demonstrate that PWPL enables coherent control of spin exchange in ion-atom collisions, far outside the ultracold regime, even with as many as 5000 partial wave contributions. The predicted extent of control is sufficient to be measurable in cold atom-ion hybrid experiments.
\end{abstract}

\date{\today}
\maketitle
{\it Introduction--} 
Two-body collisions and chemical reactions of atoms and molecules are responsible for a wide range of phenomena in physics and chemistry\cite{Levine_book,May_Kuhn_book,Nitzan_book}, such as energy transfer, thermalization, relaxation, decoherence, and spectral line broadening to name a few. As these phenomena determine the properties of dilute gases, they play a central role in atomic and molecular spectroscopy \cite{Hartmann_book}, atmospheric science \cite{Massey_book}, astrochemistry \cite{Smith2011}, and ultracold chemistry \cite{Krems_book,Dulieu_book}. For this reason, controlling the quantum dynamics of two-body collisions has long been a major thrust of physics and chemistry, and led to the development of several vast fields of research, including coherent control \cite{Brumer_book}, laser control of chemical reactions \cite{Rice_book}, mode-selective chemistry \cite{Levine_book} and stereochemistry \cite{Alexander1998,Orr-Erwing1996}.

A key challenge in controlling binary collisions and chemical reactions lies in the random nature of scattering events. Specifically, under ambient conditions, the integral collision cross section (or reaction rate) is determined by many partial wave contributions, which are essentially random functions of  $\ell$, the orbital angular momentum for the collision \cite{Mott_book,Levine_book}. Quantum control protocols target a given partial wave contribution and rely in the phase of the underlying scattering amplitude (or S-matrix). Thus, the optimal values of the control parameter (be it the value of an external field in field-based control schemes, laser pulse parameters in optimal control, or superposition parameters in coherent control) are necessarily $\ell$-dependent, and thus cannot be optimized for all values of $\ell$ contributing to the integral scattering cross section. This fundamental issue, which we will refer to as "partial wave scrambling" has generally prevented the application of quantum control techniques to collisions \cite{Krems2008} and the observation of scattering resonances \cite{Skodje2000} in the multiple partial wave regime.

A common way to combat partial wave scrambling is to cool the colliding species down to the ultracold regime, where collisions are dominated by a single initial partial wave. This eliminates the scrambling in the incident collision channel and reduces its severity in the outgoing channels, allowing for a high degree of control. The current state-of-art control techniques rely on cold molecules/atoms prepared in a well-defined internal state interacting with electric \cite{Wang2015, Li2021, Quemener2022}, microwave \cite{Karman2018, Lassabliere2018, Anderegg2021,Schindewolf2022,Chen2023}, optical \cite{Xie2020}, or magnetic fields \cite{Yang2019, Tscherbul2020}. However, the use of external fields may not be suitable for some applications, especially when the collision partners lack electric and/or magnetic dipole moments \cite{Devolder2021}.

Coherent control is an attractive method that does not rely on external fields. This technique involves preparing superpositions of the internal states of colliding particles to create interference effects that can be manipulated by changing the relative phase between the states \cite{Brumer_book,Shapiro1996,Gong2003}. Complete coherent control is possible over ultracold resonant exchange processes, such as spin, charge, or excitation exchange, where only a single partial wave is involved in both the incident and final collision channels \cite{Devolder2021}. These processes can be completely suppressed (or activated), via destructive (constructive) interference. By contrast, coherent control in the multiple partial wave regime can face a major challenge due to the partial wave scrambling.

Here, we show that  efficient coherent control in the multiple partial wave regime can be achieved using the partial wave phase locking (PWPL) effect \cite{Sikorsky2018,Kartz2022,Tomza2019,Cote2018}, which manifests in a coherent addition of different partial wave contribution. The physical origin of the PWPL effect can be attributed to the short-range nature of the spin-exchange interaction and the small magnitude of the centrifugal kinetic energy compared to the well depth of the interaction potential \cite{Gao2001},  enhancing quantum interference in the multiple partial wave regime. Here, we show this robust PWPL-assisted coherent control of spin exchange in ion-atom collisions (Sr$^+$-Rb). Cold atom-ion hybrid systems have been realized experimentally as a promising platform for quantum science \cite{Tomza2019}. Hence our results can be readily verified in the laboratory. To our knowledge, this is the first approach to control collisions beyond the ultracold regime and can be applied to a wide range of quasiresonant process \cite{Cote2018}.

{\it Initial superposition and coherent control of cross section--} Consider a binary collision $A+B$, where $A$ and $B$ denote atoms or molecules initially prepared in a coherent superposition of internal angular momentum states :
\begin{equation}
	\ket{\psi_{A}}=N\left(\sqrt{\cos\eta}\ket{j_A,m_{1A}}+\sqrt{\sin\eta}e^{i\frac{\beta}{2}}\ket{j_A,m_{2A}}\right),
\end{equation}
\begin{equation}
	\ket{\psi_{B}}=N\left(\sqrt{\sin\eta}e^{i\frac{\beta}{2}}\ket{j_B,m_{1B}}+\sqrt{\cos\eta}\ket{j_B,m_{2B}}\right),
\end{equation}
where $N=\frac{1}{\sqrt{\sin\eta+\cos\eta}}$ is a normalization factor, $\eta\in[0,\pi/2]$ and $\beta\in[0,2\pi]$ are the parameters that determine the relative population and phase of the superposition, $j_A$ and $j_B$ are the internal angular momenta of the colliding partners $A$ and $B$, and $m_{1A}$, $m_{2A}$, $m_{1B}$, and $m_{2B}$ are the corresponding projections on the space-fixed quantization axis $Z$, subject to the constraint $m_{1A}+m_{2B}=m_{1B}+m_{2A}$ imposed by rotational symmetry, which is required to obtain interference \cite{Omiste2018}. The initial state for the collision is given by the product $\ket{\psi_{A}}\ket{\psi_{B}}$:
\begin{equation}
	\begin{split}
		\ket{\Psi_{sup}}=N^2\Bigg(\cos\eta \ket{m_{1A};m_{2B}}+\sin\eta e^{i\beta}\ket{m_{2A};m_{1B}}\\+\sqrt{\cos\eta\sin\eta}e^{i\frac{\beta}{2}}(\ket{m_{1A};m_{1B}}+\ket{m_{2A};m_{2B}})\Bigg),
	\end{split}
	\label{eq:psi_nent}
\end{equation}
where we have defined $\ket{m_{1A};m_{2B}}\equiv\ket{j_A,m_{1A}}\ket{j_B,m_{2A}}$ etc, for brevity. Due to rotational symmetry, the states $\ket{m_{1A};m_{2B}}$ and $\ket{m_{2A};m_{1B}}$ interfere with each other, while the states $\ket{m_{1A};m_{1B}}$ and $\ket{m_{2A};m_{2B}}$ instead give rise to satellite terms\cite{Brumer_book,Devolder2021}. For this reason, the superposition,
\begin{equation}
	\ket{\Psi_{ent}}=\cos\eta \ket{m_{1A};m_{2B}}+\sin\eta e^{i\beta}\ket{m_{2A};m_{1B}},
	\label{eq:psi_ent}
\end{equation}
provides better control, as previously demonstrated \cite{Devolder2022}. While this superposition is harder to prepare experimentally than that in eq. (\ref*{eq:psi_nent}) because the colliding partners are entangled, it provides a useful reference point for comparing different control schemes, as shown below.

The cross section from the non-entangled superposition (\ref{eq:psi_nent}) to a final state $\ket{f}$ can be split into two parts: one related to the cross section from the entangled superposition (\ref{eq:psi_ent}), and one from the satellite terms:
\begin{equation}
	\sigma_{sup\rightarrow f}(\eta,\beta) =N^4 \left(\sigma_{ent\rightarrow f}+\sigma_{sat\rightarrow f}\right),
\end{equation}
where 
\begin{equation}
	\begin{split}
		\sigma_{ent\rightarrow f}(\eta,\beta)=\frac{\pi}{k^2}\sum_{\ell,m_\ell}\sum_{\ell',m'_\ell}\Big|&\cos\eta \ T_{m_{1A}m_{2B}\ell m_\ell\rightarrow f\ell'm'_\ell}\\+&\sin\eta\ T_{m_{2A}m_{1B}\ell m_\ell\rightarrow f\ell'm'_\ell}\Big|^2,
	\end{split}
	\label{eq:sig_ent}
\end{equation}
and 
\begin{equation}
	\sigma_{sat\rightarrow f}(\eta)=\cos\eta\sin\eta\left(\sigma_{m_{1A};m_{1B}\rightarrow f}+\sigma_{m_{2A};m_{2B}\rightarrow f}\right).
\end{equation}
Here, $\ell$ and $\ell'$ are the initial and final orbital angular momenta for the collision, while $m_\ell$ and $m'_\ell$ are the projections of $\ell$ and $\ell'$ on a space-fixed quantization axis $Z$, $T_{m_{1A}m_{2B}\ell m_\ell\rightarrow f\ell'm'_\ell}$ and $T_{m_{2A}m_{1B}\ell m_\ell\rightarrow f\ell'm'_\ell}$ are the $T$-matrix elements associated with the initial states $\ket{m_{1A};m_{2B}}$ and $\ket{m_{2A};m_{1B}}$, and $k$ is the relative wavevector.

To achieve good control of $\sigma_{sup\rightarrow f}$, it is necessary to exert efficient control over $\sigma_{ent\rightarrow f}$. However, even with good control over the latter, the inclusion of large satellite terms can significantly reduce overall control. This illustrates two important requirements for the efficient coherent control of $\sigma_{sup\rightarrow f}$: (i) achieving the best possible control over $\sigma_{ent\rightarrow f}$ and (ii) minimizing the value of $\sigma_{sat\rightarrow f}$. To quantify these requirements, we define two control indices. First, the extent of control over the cross section from the entangled initial superposition $\sigma_{ent\rightarrow f}$ can be defined as:
\begin{equation}
	R_{c,ent}=\frac{\left|\sigma_{int}\right|}{\sqrt{\sigma_{m_{1A};m_{2B}\rightarrow f}\sigma_{m_{2A};m_{1B}\rightarrow f}}},
\end{equation}
where
\begin{equation}
	\sigma_{int}=\frac{\pi}{k^2}\sum_{\ell,m_\ell}\sum_{\ell',m'_\ell}T_{m_{1A}m_{2B}\ell m_\ell\rightarrow f\ell'm'_\ell}T_{m_{2A}m_{1B}\ell m_\ell\rightarrow f\ell'm'_\ell}^*,
	\label{eq:cross_int}
\end{equation}
is the interference contribution to the integral cross section, and $\sigma_{m_{1A};m_{1B}\rightarrow f}$ and $\sigma_{m_{2A};m_{2B}\rightarrow f}$ are the cross sections from the states $\ket{m_{1A};m_{1B}}$ and $\ket{m_{2A};m_{2B}}$ respectively, to the final state $\ket{f}$. The definition relies on the Schwartz inequality, so that $R_{c,ent}$ lies between zero and one. Second, the effect of the satellite terms is quantified by:
\begin{equation}
	R_{sat}=\frac{min\left(\sigma_{m_{1A};m_{2B}\rightarrow f},\sigma_{m_{2A};m_{1B}\rightarrow f}\right)}{max\left(\sigma_{m_{1A};m_{1B}\rightarrow f},\sigma_{m_{2A};m_{2B}\rightarrow f}\right)}.
	\label{eq.R_sat}
\end{equation} 
Here, $min\left(\sigma_{m_{1A};m_{2B}\rightarrow f},\sigma_{m_{2A};m_{1B}\rightarrow f}\right)$ and ($max\left(\sigma_{m_{1A};m_{1B}\rightarrow f},\sigma_{m_{2A};m_{2B}\rightarrow f}\right)$) is the smallest (largest) of the two values. Small satellite terms correspond to $R_{sat}>>1$, a favorable condition for coherent control.

Finally, we define a global control index as $R_{c,sup}=max_\eta(V)$, where $V$ represents the visibility that measures the oscillation of $\sigma_{sup\rightarrow f}(\eta,\beta)$ when only the relative phase $\beta$ is varied: 
\begin{equation}
	V(\tilde{\eta})=\frac{\sigma_{sup \rightarrow f}(\tilde{\eta},\beta_{max}^{\tilde{\eta}})-\sigma_{sup \rightarrow f}(\tilde{\eta},\beta_{min}^{\tilde{\eta}})}{\sigma_{sup \rightarrow f}(\tilde{\eta},\beta_{max}^{\tilde{\eta}})+\sigma_{sup \rightarrow f}(\tilde{\eta},\beta_{min}^{\tilde{\eta}})},
\end{equation}
where $\beta_{min}^{\tilde{\eta}}$ ($\beta_{max}^{\tilde{\eta}})$ is the value of $\beta$ for which the cross-section is minimal (maximal) when $\eta=\tilde{\eta}$. Like $R_{c,ent}$, the values of $R_{c,sup}$ are bound between 0 and 1.

{\it Partial wave scrambling in coherent control-}
The control of the entangled cross section $\sigma_{ent\rightarrow f}$ is limited by incoherent addition of the initial and final partial waves $(\ell m_\ell,\ell' m'_\ell)$ in Eq. (\ref{eq:sig_ent}). For a given superposition determined by the parameters ($\eta$ and $\beta$), some partial wave contributions $(\ell m_\ell,\ell' m'_\ell)$ may experience constructive interference while others may exhibit destructive interference, suppressing the interference term in Eq. (\ref{eq:cross_int}) \cite{Devolder2023}. This partial wave scrambling issue becomes more significant as the collision energy increases, along with the number of $(\ell m_\ell,\ell' m'_\ell)$ contributions, and can result in complete loss of control.

To quantify partial wave scrambling, we examine the distribution of the superposition parameters ($\eta,\beta$) for which each partial wave contribution reaches a minimum value, $\eta_{min}^{\ell m_\ell,\ell' m'_\ell}=\arctan\left(\frac{\left|T_{m_{1A}m_{2B}\ell m_\ell\rightarrow f\ell'm'_\ell}\right|}{\left|T_{m_{2A}m_{1B}\ell m_\ell\rightarrow f\ell'm'_\ell}\right|}\right)$ and $\beta_{min}^{\ell m_\ell,\ell' m'_\ell}=arg\left(T_{m_{1A}m_{2B}\ell m_\ell\rightarrow f\ell',m'_\ell}\right)-arg\left(T_{m_{2A}m_{1B}\ell m_\ell\rightarrow f\ell',m'_\ell}\right)$  depending on the ratio of the magnitudes, and on the difference of phases (arguments) of the $T$-matrix elements, respectively. Note that the maximum parameters are simply obtained from: $\eta_{min}^{\ell m_\ell,\ell' m'_\ell}+\eta_{max}^{\ell,m_\ell,\ell'm'_\ell}=\pi/2$ and $\beta_{max}^{\ell m_\ell,\ell' m'_\ell}-\beta_{min}^{\ell m_\ell,\ell' m'_\ell}=\pi$.   The distribution of the optimal parameters $\eta_{min}^{\ell m_\ell,\ell' m'_\ell}$ and $\beta_{min}^{\ell m_\ell,\ell' m'_\ell}$ determines the degree of partial wave scrambling. A random distribution leads to a rapid decrease in control as the number $N_{\ell}$ of significant partial wave contributions $(\ell m_\ell,\ell' m'_\ell)$ increases, because $R_{c,ent}$ scales as $1/\sqrt{N_{\ell m_\ell,\ell' m'_\ell}}$ \cite{Devolder2023}. The solution of the partial wave scrambling problem lies in the clustering of the optimal parameters $\eta_{min}^{\ell m_\ell,\ell' m'_\ell}$ and $\beta_{min}^{\ell m_\ell,\ell' m'_\ell}$. As shown below, the clustering of $\beta_{min}^{\ell m_\ell,\ell' m'_\ell}$ is manifest in the PWPL mechanism, which dramatically reduces the partial wave scrambling and thereby paves the way to coherent control in the multiple partial wave regime. 

{\it Application: Coherent control of spin relaxation in Sr$^+$-Rb collisions--}
The PWPL phenomenon was first predicted to occur in spin relaxation in ion-atom collisions, making hybrid ion-atom systems \cite{Tomza2019} ideal for investigating PWPL assisted coherent control. Consider a rubidium atom ($^{87}$Rb) prepared in a superposition of hyperfine states $\ket{2,-1}_B$ and $\ket{2,0}_B$, colliding with a trapped strontium ion $^{88}$Sr$^+$ prepared in a superposition of Zeeman states $\ket{1/2,-1/2}_A$ and $\ket{1/2,1/2}_A$. We present the results for relaxation to the final states $\ket{1/2,-1/2}_A \ket{1,0}_B\equiv \ket{\downarrow} $ and $\ket{1/2,+1/2}_A \ket{1,0}_B\equiv \ket{\uparrow} $, as these states have larger cross-sections than the final states $\ket{1/2,\pm1/2}_A \ket{1,-1}_B$ and $\ket{1/2,\pm1/2}_A \ket{1,1}_B$. 

To motivate experimental studies we carried out rigorous coupled-channel (CC) calculations of Sr$^+$-Rb collisions using state of the art ab-initio interaction potentials and second-order spin-orbit interactions, as described in \cite{Sikorsky2018}. To ensure numerical convergence of the results for collision energies ranging from 1 $\mu$K to 50 mK, we used extended CC basis sets including up to 80 partial waves.
\begin{figure}[t]
	\centering
	\includegraphics[width=\columnwidth]{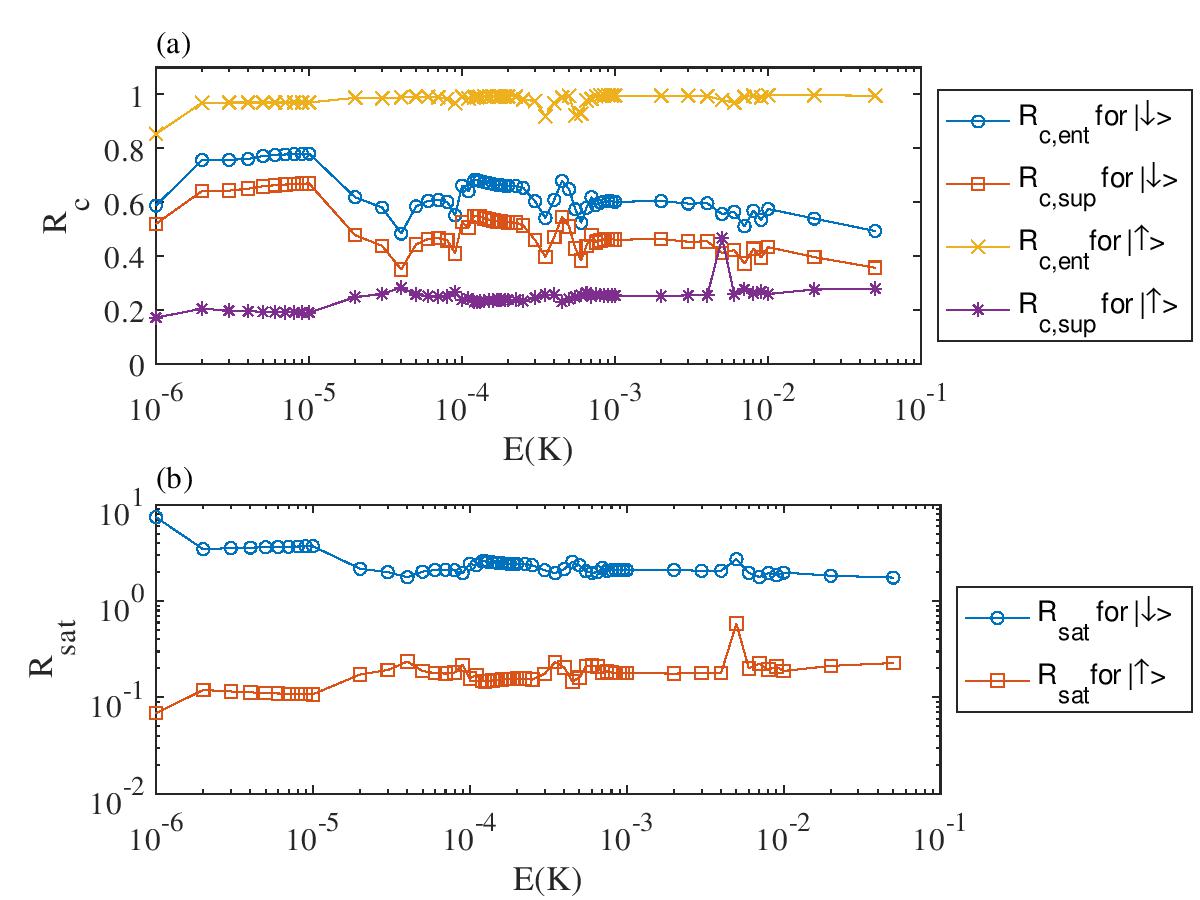}
	\caption{(a) Control indices, $R_{c,ent}$ and $R_{c,sup}$, for the transition from the entangled and non-entangled superpositions to the final states $\ket{\downarrow}$ and $\ket{\uparrow}$ . (b) Satellite term index $R_{sat}$ for the final states $\ket{\downarrow}$ and $\ket{\uparrow}$}
	\label{fig:Rc_sat}
\end{figure}

The control indices, $R_{c,ent}$ and $R_{c,sup}$, for Sr$^+$-Rb calculated from exact CC results, are shown in Fig. \ref{fig:Rc_sat} (a). The high value of the entangled control index, $R_{c,ent}$, demonstrates that efficient control is possible in the multiple partial-wave regime. For the final state $\ket{\uparrow}$, $R_{c,ent}$ is close to 1, indicating complete control, whereas for the final state $\ket{\downarrow}$, $R_{c,ent}$ is around 0.5-0.6. Remarkably, the high entangled control index remains mostly independent of collision energy, indicating robust control over a wide energy range. Variation of the control can be caused by the presence of resonances, for example at $E$= 100 $\mu$K, 400 $\mu$K, and 5 mK.

As expected, the control index for the non-entangled superposition, $R_{c,sup}$, is affected by satellite terms, which can be quantified using the parameters $R_{sat}$ (eq. \ref{eq.R_sat}) shown in Fig. \ref{fig:Rc_sat} (b). For the final state $\ket{\uparrow}$, the satellite terms are significant, resulting in a large difference between $R_{c,ent}$ and $R_{c,sup}$, with $R_{c,sup}$ being around 0.2-0.3. In contrast, the satellite terms are small for the final state $\ket{\downarrow}$, resulting in a small difference between $R_{c,ent}$ and $R_{c,sup}$, with $R_{c,sup} \approx$ 0.4-0.5. The reason for the smaller impact of the satellite terms on the final state $\ket{\downarrow}$ is that for this state, the interfering transitions conserve the total internal angular momentum projection, $m_A^i+m_B^i=m_A^f+m_B^f$, whereas the transitions in the satellite terms do not. The situation is the opposite for the final state $\ket{\uparrow}$. These results highlight a complex trade-off between the partial wave scrambling and satellite terms. Even though partial wave scrambling is less significant for the final state $\ket{\uparrow}$, the overall control is better for the final state $\ket{\downarrow}$ due to the smaller effect of the satellite terms. 
\begin{figure}[t]
	\centering
	\includegraphics[width=\columnwidth]{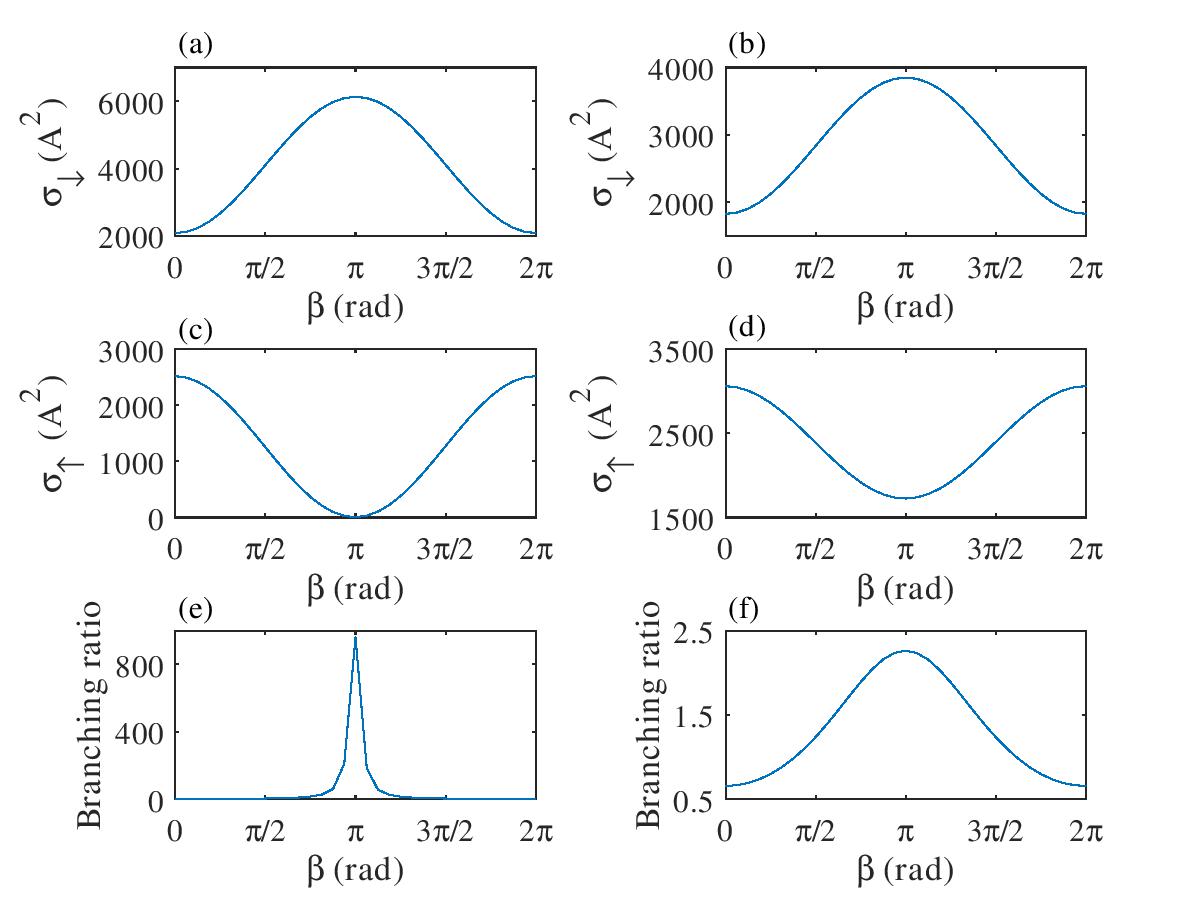}
	\caption{Coherent control of scattering cross sections and branching ratios for Sr$^+$-Rb collisions with 5000 partial wave contributions $\left(\ell m_\ell,\ell' m'_\ell\right)$. Cross sections (a) from $\ket{\Psi_{ent}}$ to $\ket{\downarrow}$, (b) from $\ket{\Psi_{sup}}$ to $\ket{\downarrow}$, (c) from $\ket{\Psi_{ent}}$ to $\ket{\uparrow}$ and (d) from $\ket{\Psi_{sup}}$ to $\ket{\uparrow}$. Branching ratios $\sigma_\downarrow/\sigma_\uparrow$ (e) from $\ket{\Psi_{ent}}$ and (f) from $\ket{\Psi_{sup}}$. The parameter $\eta$ is kept constant, $\eta=\pi/2$ for $\sigma_{\downarrow}$, while $\eta=21\pi/32$ for $\sigma_{\uparrow}$ and the branching ratios.}
	\label{fig:ctrl_beta}
\end{figure}

\begin{figure}[t]
	\centering
	\includegraphics[width=\columnwidth]{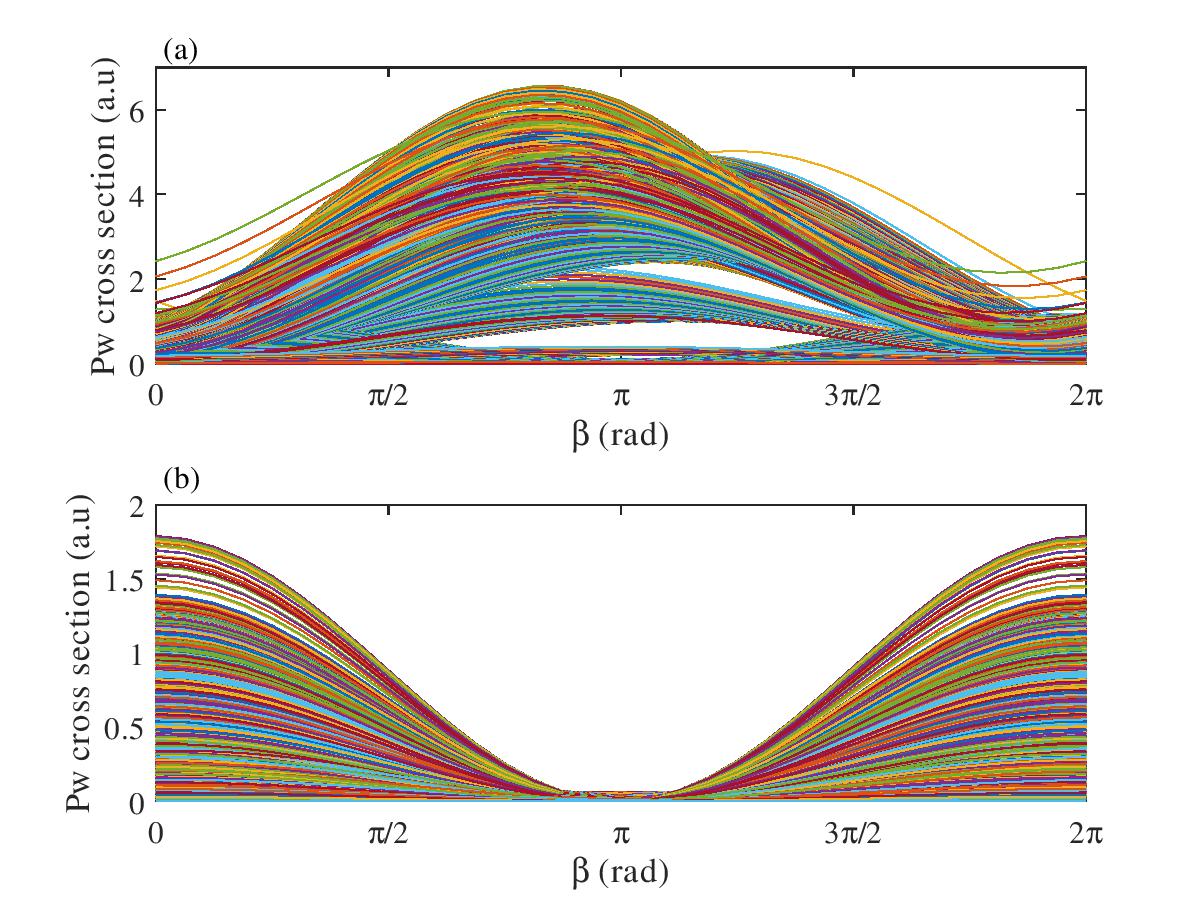}
	\caption{Phase variation of the 5000 partial wave contributions $\left(\ell m_\ell,\ell' m'_\ell\right)$ to the cross sections for the final states (a) $\ket{\downarrow}$ and (b) $\ket{\uparrow}$. The value of $\eta$ is fixed at $\eta=\pi/2$ and $21\pi/32$ for the final states $\ket{\downarrow}$ and $\ket{\uparrow}$, respectively. All partial wave contributions are plotted to illustrate the phase locking, with less efficiency observed for the final state $\ket{\downarrow}$.}
	\label{fig:phase_locking}
\end{figure}

To illustrate coherent control in the multiple partial wave regime, we consider the experimentally realistic case of Sr$^+$-Rb collisions at 50 mK \cite{Sikorsky2018}. At this collision energy, approximately 5000 $(\ell m_\ell,\ell' m'_\ell)$ partial wave states contribute to the cross sections. Figures \ref{fig:ctrl_beta}(a)-(d) illustrate coherent control of the cross section by varying the phase angle $\beta$ of the initial superposition (\ref{eq:psi_ent}). To obtain the best visibility $V$, the value of $\eta$ is fixed at $\eta=\pi/2$ and $21\pi/32$ for the final states $\ket{\downarrow}$ and $\ket{\uparrow}$, respectively. We observe a remarkable instance of complete control of scattering from the entangled superposition to the final state $\ket{\uparrow}$, showing near vanishing of the cross section due to destructive interference (see Fig. \ref{fig:ctrl_beta} (c)). The minimum value of the cross-section is 6.2 a.u., which is three orders of magnitude smaller than the maximum value of 2517.3 a.u. This complete control is made possible by the clustering of the optimal control parameters $\eta_{min}^{\ell,m_\ell,\ell',m'\ell}$ and $\beta_{min}^{\ell,m_\ell,\ell',m'\ell}$ [see Fig. \ref{fig:phase_locking}(c) and (d)], caused by PWPL. Note that if these parameters were randomly distributed, the entangled control index $R_{c,ent}$ would be equal to $\sim 1/\sqrt{5000}=0.01$, significantly smaller than the calculated value. When the initial superposition is non-entangled, complete destructive interference is countered by the presence of satellite terms,  resulting in a variation from 1926.6 to 3350.2 a.u, which remains significant and certainly large enough to be detected in modern hybrid trapped ion-atom collision experiments \cite{Sikorsky2018,Kartz2022,Tomza2019,Cote2018}. 

\begin{figure}[t]
	\centering
	\includegraphics[width=\columnwidth]{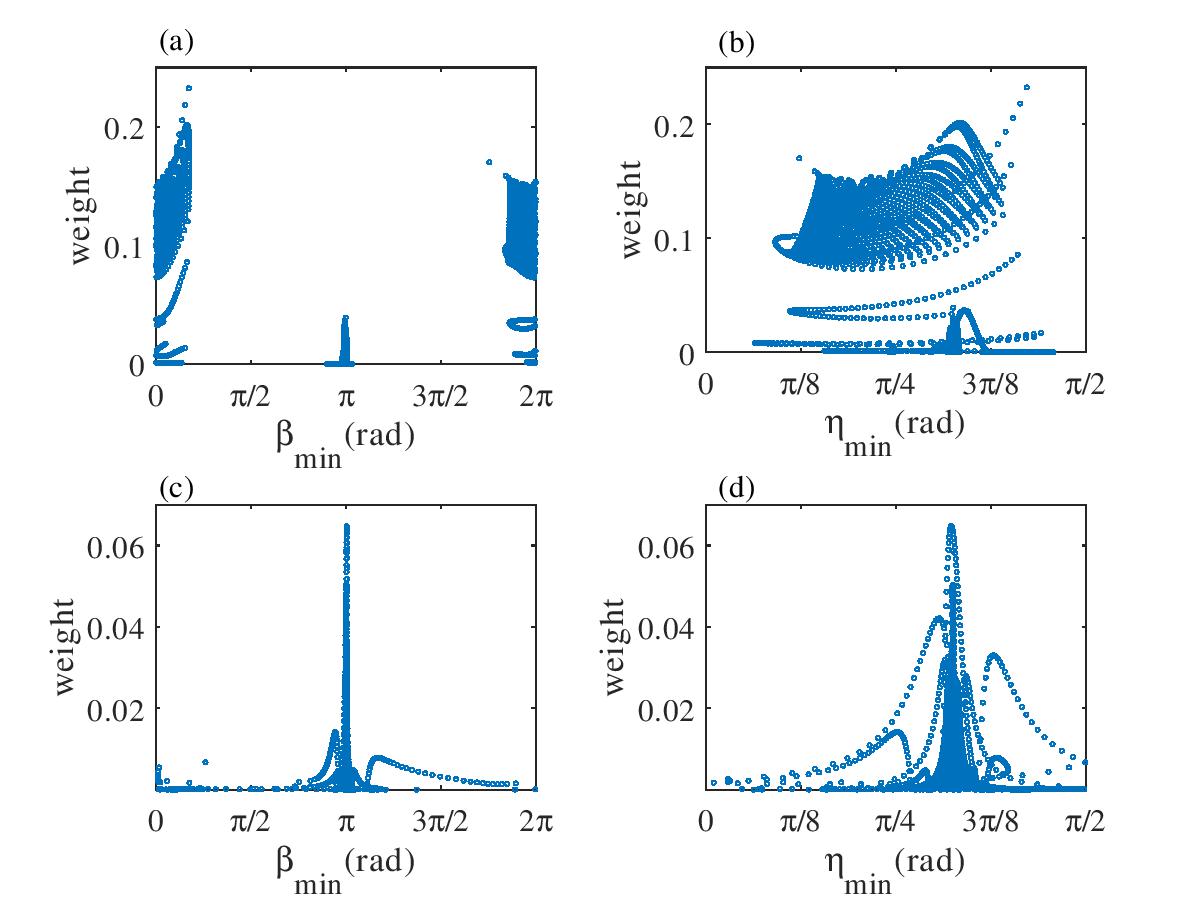}
	\caption{Distribution of the optimal parameters $\eta_{min}^{\ell m_\ell,\ell' m'_\ell}$ and $\beta_{min}^{\ell,m_\ell,\ell',m'\ell}$ for the final states ((a) and (b)) $\ket{\downarrow}$ and ((c) and (d)) $\ket{\uparrow}$. Each point corresponds to a pair $(\ell m_\ell,\ell' m'_\ell)$. The weight $w$ is calculated by the sum of squared magnitudes of T-matrix elements: $w=\left|T_{m_{1A}m_{2B}\ell m_\ell\rightarrow f\ell'm'_\ell}\right|^2+\left|T_{m_{2A}m_{1B}\ell m_\ell\rightarrow f\ell'm'_\ell}\right|^2$}
	\label{fig:dist_pw}
\end{figure}

For the final state $\ket{\downarrow}$, the  clustering of $\beta_{min}^{\ell,m_\ell,\ell',m'\ell}$ is less effective than for $\ket{\uparrow}$, as shown in Fig.\ref{fig:phase_locking} (a) and  \ref{fig:dist_pw} (a). More detrimential for the control, the distribution of the other optimal parameter, $\eta_{min}^{\ell m_\ell,\ell' m'_\ell}$ [see Fig. \ref{fig:dist_pw} (b)] is broad, highlighting the importance of the distribution of $\eta_{min}^{\ell m_\ell,\ell' m'_\ell}$. The PWPL effect locks the difference of  T-matrix phases but not the ratio of their magnitudes, and thus does not guarantee the clustering of $\eta_{min}^{\ell m_\ell,\ell' m'_\ell}$. While this limitation can prevent PWPL from completely solving the problem of partial wave scrambling, we observe that despite the relatively broad distribution of ratios, a good degree of control is still achievable in the multiple partial wave regime, with $\sigma_{sup\rightarrow f}(\eta,\beta)$ ranging from 2085.5 to 6131.7 a.u. The inclusion of the satellite terms has a slight impact on the control, resulting in a variation from 1827.7 to 3850.7 a.u.

As noted above, the control of the scattering to the final states $\ket{\downarrow}$ and $\ket{\uparrow}$ is optimized in different regions of parameter space allowing for efficient control of the corresponding branching ratio $\sigma_\downarrow/\sigma_\uparrow$, as shown in Fig. \ref{fig:ctrl_beta}(e)-(f) at $\eta=21\pi/32$. With the entangled superposition Eq. (\ref{eq:psi_ent}), extremely robust control is achieved, with $\sigma_\downarrow/\sigma_\uparrow$ varying by three orders of magnitude (from 0.94941 to 960.66), thanks to the complete control of $\sigma_\uparrow$. When using the non-entangled superposition (\ref{eq:psi_nent}), the presence of satellite terms limits the extent of control over the branching ratio to 0.66-2.26, which is large enough to be experimentally measurable.

{\it Conclusion-} In summary, we have shown that partial wave phase-locking enables coherent control in the multiple partial wave regime via a dramatic reduction of partial wave scrambling. The clustering of the optimal superposition parameters $\beta_{\text{min}}^{\ell,m_\ell,\ell',m_\ell'}$ enabled by PWPL allows for the synchronized control of different partial wave contributions to the total scattering cross section. In cases where the distribution of the optimal control parameters $\eta_{\text{min}}^{\ell,m_\ell,\ell',m_\ell'}$ is broad, such as for the final state $\ket{\downarrow}$, partial-wave scrambling is only partially eliminated by PWPL. Even though the satellite terms reduce the control with non-entangled superpositions, our rigorous CC calculations show that coherent control over state-to-state integral cross sections and of the branching ratios is significant and measurable. Therefore, collisions between $^{87}$Rb and $^{88}$Sr$^+$, observed in a series of recent experiments \cite{Sikorsky2018,Kartz2022,Tomza2019}, appear to be ideal for the first experimental observation of coherent control of two-body scattering outside of the ultracold domain. As the PWPL phenomenom was shown to apply to any quasiresonant scattering process \cite{Cote2018}, a wide range of these processes could soon become amenable to robust coherent control. Furthermore, the extreme sensitivity of coherent control to PWPL effect implies that the study of collisions of atoms and molecules prepared in coherent superposition of internal states provides an ideal approach for investigating the PWPL effect.

This work was supported by the U.S. Air Force Office of Scientific Research (AFOSR) under Contract No.FA9550-22-1-0361.

\bibliography{Phase_locking}

\begin{thebibliography}{37}%
\makeatletter
\providecommand \@ifxundefined [1]{%
 \@ifx{#1\undefined}
}%
\providecommand \@ifnum [1]{%
 \ifnum #1\expandafter \@firstoftwo
 \else \expandafter \@secondoftwo
 \fi
}%
\providecommand \@ifx [1]{%
 \ifx #1\expandafter \@firstoftwo
 \else \expandafter \@secondoftwo
 \fi
}%
\providecommand \natexlab [1]{#1}%
\providecommand \enquote  [1]{``#1''}%
\providecommand \bibnamefont  [1]{#1}%
\providecommand \bibfnamefont [1]{#1}%
\providecommand \citenamefont [1]{#1}%
\providecommand \href@noop [0]{\@secondoftwo}%
\providecommand \href [0]{\begingroup \@sanitize@url \@href}%
\providecommand \@href[1]{\@@startlink{#1}\@@href}%
\providecommand \@@href[1]{\endgroup#1\@@endlink}%
\providecommand \@sanitize@url [0]{\catcode `\\12\catcode `\$12\catcode
  `\&12\catcode `\#12\catcode `\^12\catcode `\_12\catcode `\%12\relax}%
\providecommand \@@startlink[1]{}%
\providecommand \@@endlink[0]{}%
\providecommand \url  [0]{\begingroup\@sanitize@url \@url }%
\providecommand \@url [1]{\endgroup\@href {#1}{\urlprefix }}%
\providecommand \urlprefix  [0]{URL }%
\providecommand \Eprint [0]{\href }%
\providecommand \doibase [0]{https://doi.org/}%
\providecommand \selectlanguage [0]{\@gobble}%
\providecommand \bibinfo  [0]{\@secondoftwo}%
\providecommand \bibfield  [0]{\@secondoftwo}%
\providecommand \translation [1]{[#1]}%
\providecommand \BibitemOpen [0]{}%
\providecommand \bibitemStop [0]{}%
\providecommand \bibitemNoStop [0]{.\EOS\space}%
\providecommand \EOS [0]{\spacefactor3000\relax}%
\providecommand \BibitemShut  [1]{\csname bibitem#1\endcsname}%
\let\auto@bib@innerbib\@empty
\bibitem [{\citenamefont {Levine}(2005)}]{Levine_book}%
  \BibitemOpen
  \bibfield  {author} {\bibinfo {author} {\bibfnamefont {R.~D.}\ \bibnamefont
  {Levine}},\ }\href@noop {} {\emph {\bibinfo {title} {Molecular Reaction
  Dynamics}}}\ (\bibinfo  {publisher} {Cambridge University Press},\ \bibinfo
  {year} {2005})\BibitemShut {NoStop}%
\bibitem [{\citenamefont {May}\ and\ \citenamefont
  {Kuhn}(2011)}]{May_Kuhn_book}%
  \BibitemOpen
  \bibfield  {author} {\bibinfo {author} {\bibfnamefont {V.}~\bibnamefont
  {May}}\ and\ \bibinfo {author} {\bibfnamefont {O.}~\bibnamefont {Kuhn}},\
  }\href@noop {} {\emph {\bibinfo {title} {Charge and Energy Transfer Dynamics
  in Molecular Systems}}}\ (\bibinfo  {publisher} {Wiley-VCH},\ \bibinfo {year}
  {2011})\BibitemShut {NoStop}%
\bibitem [{\citenamefont {Nitzan}(2006)}]{Nitzan_book}%
  \BibitemOpen
  \bibfield  {author} {\bibinfo {author} {\bibfnamefont {A.}~\bibnamefont
  {Nitzan}},\ }\href@noop {} {\emph {\bibinfo {title} {Chemical Dynamics in
  Condensed Phases}}}\ (\bibinfo  {publisher} {Oxford University Press},\
  \bibinfo {year} {2006})\BibitemShut {NoStop}%
\bibitem [{\citenamefont {Hartmann}\ \emph {et~al.}(2008)\citenamefont
  {Hartmann}, \citenamefont {Boulet},\ and\ \citenamefont
  {Robert}}]{Hartmann_book}%
  \BibitemOpen
  \bibfield  {author} {\bibinfo {author} {\bibfnamefont {J.-M.}\ \bibnamefont
  {Hartmann}}, \bibinfo {author} {\bibfnamefont {C.}~\bibnamefont {Boulet}},\
  and\ \bibinfo {author} {\bibfnamefont {D.}~\bibnamefont {Robert}},\
  }\href@noop {} {\emph {\bibinfo {title} {Collisional Effects on Molecular
  Spectra}}}\ (\bibinfo  {publisher} {Elsevier},\ \bibinfo {year}
  {2008})\BibitemShut {NoStop}%
\bibitem [{\citenamefont {Massey}\ and\ \citenamefont
  {Bates}(1982)}]{Massey_book}%
  \BibitemOpen
  \bibfield  {author} {\bibinfo {author} {\bibfnamefont {H.~S.~W.}\
  \bibnamefont {Massey}}\ and\ \bibinfo {author} {\bibfnamefont {D.~R.}\
  \bibnamefont {Bates}},\ }\href@noop {} {\emph {\bibinfo {title} {Applied
  Atomic Collision physics: Atmospheric Physics and Chemistry}}}\ (\bibinfo
  {publisher} {Academic Press},\ \bibinfo {year} {1982})\BibitemShut {NoStop}%
\bibitem [{\citenamefont {Smith}(2011)}]{Smith2011}%
  \BibitemOpen
  \bibfield  {author} {\bibinfo {author} {\bibfnamefont {I.~W.~M.}\
  \bibnamefont {Smith}},\ }\bibfield  {title} {\bibinfo {title} {Laboratory
  astrochemistry: Gas-phase processes},\ }\href@noop {} {\bibfield  {journal}
  {\bibinfo  {journal} {Annu. Rev. Astron. Astrophus.}\ }\textbf {\bibinfo
  {volume} {49}},\ \bibinfo {pages} {29} (\bibinfo {year} {2011})}\BibitemShut
  {NoStop}%
\bibitem [{\citenamefont {Krems}(2019)}]{Krems_book}%
  \BibitemOpen
  \bibfield  {author} {\bibinfo {author} {\bibfnamefont {R.}~\bibnamefont
  {Krems}},\ }\href@noop {} {\emph {\bibinfo {title} {Molecules in
  Electromagnetic Fields}}}\ (\bibinfo  {publisher} {Wiley},\ \bibinfo {year}
  {2019})\BibitemShut {NoStop}%
\bibitem [{\citenamefont {Dulieu}\ and\ \citenamefont
  {Osterwalder}(2017)}]{Dulieu_book}%
  \BibitemOpen
  \bibfield  {author} {\bibinfo {author} {\bibfnamefont {O.}~\bibnamefont
  {Dulieu}}\ and\ \bibinfo {author} {\bibfnamefont {A.}~\bibnamefont
  {Osterwalder}},\ }\href@noop {} {\emph {\bibinfo {title} {Cold Chemistry:
  Molecular Scattering and Reactivity Near Absolute Zero}}}\ (\bibinfo
  {publisher} {Royal Society of Chemistry},\ \bibinfo {year}
  {2017})\BibitemShut {NoStop}%
\bibitem [{\citenamefont {Shapiro}\ and\ \citenamefont
  {Brumer}(2012)}]{Brumer_book}%
  \BibitemOpen
  \bibfield  {author} {\bibinfo {author} {\bibfnamefont {M.}~\bibnamefont
  {Shapiro}}\ and\ \bibinfo {author} {\bibfnamefont {P.}~\bibnamefont
  {Brumer}},\ }\href@noop {} {\emph {\bibinfo {title} {Quantum Control of
  Molecular Processes}}}\ (\bibinfo  {publisher} {Wiley-VCH},\ \bibinfo
  {address} {Weinheim, Germany},\ \bibinfo {year} {2012})\BibitemShut {NoStop}%
\bibitem [{\citenamefont {Rice}\ and\ \citenamefont {Zhao}(2000)}]{Rice_book}%
  \BibitemOpen
  \bibfield  {author} {\bibinfo {author} {\bibfnamefont {S.}~\bibnamefont
  {Rice}}\ and\ \bibinfo {author} {\bibfnamefont {M.}~\bibnamefont {Zhao}},\
  }\href@noop {} {\emph {\bibinfo {title} {Optical Control of Molecular
  Dynamics}}}\ (\bibinfo  {publisher} {Wiley},\ \bibinfo {year}
  {2000})\BibitemShut {NoStop}%
\bibitem [{\citenamefont {Alexander}\ \emph {et~al.}(2011)\citenamefont
  {Alexander}, \citenamefont {Brouard}, \citenamefont {Kalogerakis},\ and\
  \citenamefont {Simons}}]{Alexander1998}%
  \BibitemOpen
  \bibfield  {author} {\bibinfo {author} {\bibfnamefont {A.~J.}\ \bibnamefont
  {Alexander}}, \bibinfo {author} {\bibfnamefont {M.}~\bibnamefont {Brouard}},
  \bibinfo {author} {\bibfnamefont {K.}~\bibnamefont {Kalogerakis}},\ and\
  \bibinfo {author} {\bibfnamefont {J.~P.}\ \bibnamefont {Simons}},\ }\bibfield
   {title} {\bibinfo {title} {Chemistry with a sense of direction—the
  stereodynamics of bimolecular reactions},\ }\href@noop {} {\bibfield
  {journal} {\bibinfo  {journal} {Chem. Soc. Rev.}\ }\textbf {\bibinfo {volume}
  {27}},\ \bibinfo {pages} {405} (\bibinfo {year} {2011})}\BibitemShut
  {NoStop}%
\bibitem [{\citenamefont {Orr-Erwing}(1996)}]{Orr-Erwing1996}%
  \BibitemOpen
  \bibfield  {author} {\bibinfo {author} {\bibfnamefont {A.~J.}\ \bibnamefont
  {Orr-Erwing}},\ }\bibfield  {title} {\bibinfo {title} {Dynamical
  stereochemistry of bimolecular reactions},\ }\href@noop {} {\bibfield
  {journal} {\bibinfo  {journal} {J. Chem. Soc.,Faraday Trans.}\ }\textbf
  {\bibinfo {volume} {92}},\ \bibinfo {pages} {881} (\bibinfo {year}
  {1996})}\BibitemShut {NoStop}%
\bibitem [{\citenamefont {Mott}\ and\ \citenamefont
  {Massey}(1965)}]{Mott_book}%
  \BibitemOpen
  \bibfield  {author} {\bibinfo {author} {\bibfnamefont {N.~F.}\ \bibnamefont
  {Mott}}\ and\ \bibinfo {author} {\bibfnamefont {H.~S.~W.}\ \bibnamefont
  {Massey}},\ }\href@noop {} {\emph {\bibinfo {title} {The Theory of Atomic
  Collisions}}}\ (\bibinfo  {publisher} {Oxford University Press},\ \bibinfo
  {year} {1965})\BibitemShut {NoStop}%
\bibitem [{\citenamefont {Krems}(2008)}]{Krems2008}%
  \BibitemOpen
  \bibfield  {author} {\bibinfo {author} {\bibfnamefont {R.}~\bibnamefont
  {Krems}},\ }\bibfield  {title} {\bibinfo {title} {Cold controlled
  chemistry},\ }\href@noop {} {\bibfield  {journal} {\bibinfo  {journal} {Phys.
  Chem. Chem. Phys.}\ }\textbf {\bibinfo {volume} {10}},\ \bibinfo {pages}
  {4079} (\bibinfo {year} {2008})}\BibitemShut {NoStop}%
\bibitem [{\citenamefont {Skodje}\ \emph {et~al.}(2000)\citenamefont {Skodje},
  \citenamefont {Skouteris}, \citenamefont {Manolopoulos}, \citenamefont {Lee},
  \citenamefont {Dong},\ and\ \citenamefont {Liu}}]{Skodje2000}%
  \BibitemOpen
  \bibfield  {author} {\bibinfo {author} {\bibfnamefont {R.~T.}\ \bibnamefont
  {Skodje}}, \bibinfo {author} {\bibfnamefont {D.}~\bibnamefont {Skouteris}},
  \bibinfo {author} {\bibfnamefont {D.}~\bibnamefont {Manolopoulos}}, \bibinfo
  {author} {\bibfnamefont {S.-H.}\ \bibnamefont {Lee}}, \bibinfo {author}
  {\bibfnamefont {F.}~\bibnamefont {Dong}},\ and\ \bibinfo {author}
  {\bibfnamefont {K.}~\bibnamefont {Liu}},\ }\bibfield  {title} {\bibinfo
  {title} {Observation of a transition state resonance in the integral cross
  section of the {F+HD} reaction},\ }\href@noop {} {\bibfield  {journal}
  {\bibinfo  {journal} {J. Chem. Phys.}\ }\textbf {\bibinfo {volume} {112}},\
  \bibinfo {pages} {4536} (\bibinfo {year} {2000})}\BibitemShut {NoStop}%
\bibitem [{\citenamefont {Wang}\ and\ \citenamefont
  {Quéméner}(2015)}]{Wang2015}%
  \BibitemOpen
  \bibfield  {author} {\bibinfo {author} {\bibfnamefont {G.}~\bibnamefont
  {Wang}}\ and\ \bibinfo {author} {\bibfnamefont {G.}~\bibnamefont
  {Quéméner}},\ }\bibfield  {title} {\bibinfo {title} {Tuning ultracold
  collisions of excited rotational dipolar molecules},\ }\href@noop {}
  {\bibfield  {journal} {\bibinfo  {journal} {New J. Phys.}\ }\textbf {\bibinfo
  {volume} {17}},\ \bibinfo {pages} {035015} (\bibinfo {year}
  {2015})}\BibitemShut {NoStop}%
\bibitem [{\citenamefont {Li}\ \emph {et~al.}(2021)\citenamefont {Li},
  \citenamefont {Tobias}, \citenamefont {Matsuda}, \citenamefont {Miller},
  \citenamefont {Valtolina}, \citenamefont {Marco}, \citenamefont {Wang},
  \citenamefont {Lassablière}, \citenamefont {Quéméner}, \citenamefont
  {Bohn},\ and\ \citenamefont {Ye}}]{Li2021}%
  \BibitemOpen
  \bibfield  {author} {\bibinfo {author} {\bibfnamefont {J.-R.}\ \bibnamefont
  {Li}}, \bibinfo {author} {\bibfnamefont {W.~G.}\ \bibnamefont {Tobias}},
  \bibinfo {author} {\bibfnamefont {K.}~\bibnamefont {Matsuda}}, \bibinfo
  {author} {\bibfnamefont {C.}~\bibnamefont {Miller}}, \bibinfo {author}
  {\bibfnamefont {G.}~\bibnamefont {Valtolina}}, \bibinfo {author}
  {\bibfnamefont {L.~D.}\ \bibnamefont {Marco}}, \bibinfo {author}
  {\bibfnamefont {R.~R.~W.}\ \bibnamefont {Wang}}, \bibinfo {author}
  {\bibfnamefont {L.}~\bibnamefont {Lassablière}}, \bibinfo {author}
  {\bibfnamefont {G.}~\bibnamefont {Quéméner}}, \bibinfo {author}
  {\bibfnamefont {J.~L.}\ \bibnamefont {Bohn}},\ and\ \bibinfo {author}
  {\bibfnamefont {J.}~\bibnamefont {Ye}},\ }\bibfield  {title} {\bibinfo
  {title} {Tuning of dipolar interactions and evaporative cooling in a three
  dimensional molecular quantum gas},\ }\href@noop {} {\bibfield  {journal}
  {\bibinfo  {journal} {Nat. Phys.}\ }\textbf {\bibinfo {volume} {17}},\
  \bibinfo {pages} {1144} (\bibinfo {year} {2021})}\BibitemShut {NoStop}%
\bibitem [{\citenamefont {Quéméner}\ \emph {et~al.}(2022)\citenamefont
  {Quéméner}, \citenamefont {Croft},\ and\ \citenamefont
  {Bohn}}]{Quemener2022}%
  \BibitemOpen
  \bibfield  {author} {\bibinfo {author} {\bibfnamefont {G.}~\bibnamefont
  {Quéméner}}, \bibinfo {author} {\bibfnamefont {J.~F.~E.}\ \bibnamefont
  {Croft}},\ and\ \bibinfo {author} {\bibfnamefont {J.~L.}\ \bibnamefont
  {Bohn}},\ }\bibfield  {title} {\bibinfo {title} {Electric field dependence of
  complex-dominated ultracold molecular collisions},\ }\href@noop {} {\bibfield
   {journal} {\bibinfo  {journal} {Phys. Rev. A}\ }\textbf {\bibinfo {volume}
  {105}},\ \bibinfo {pages} {013310} (\bibinfo {year} {2022})}\BibitemShut
  {NoStop}%
\bibitem [{\citenamefont {Karman}\ and\ \citenamefont
  {Hutson}(2018)}]{Karman2018}%
  \BibitemOpen
  \bibfield  {author} {\bibinfo {author} {\bibfnamefont {T.}~\bibnamefont
  {Karman}}\ and\ \bibinfo {author} {\bibfnamefont {J.~M.}\ \bibnamefont
  {Hutson}},\ }\bibfield  {title} {\bibinfo {title} {Microwave shielding of
  ultracold polar molecules},\ }\href@noop {} {\bibfield  {journal} {\bibinfo
  {journal} {Phys. Rev. Lett.}\ }\textbf {\bibinfo {volume} {121}},\ \bibinfo
  {pages} {163401} (\bibinfo {year} {2018})}\BibitemShut {NoStop}%
\bibitem [{\citenamefont {Lassablière}\ and\ \citenamefont
  {Quéméner}(2018)}]{Lassabliere2018}%
  \BibitemOpen
  \bibfield  {author} {\bibinfo {author} {\bibfnamefont {L.}~\bibnamefont
  {Lassablière}}\ and\ \bibinfo {author} {\bibfnamefont {G.}~\bibnamefont
  {Quéméner}},\ }\bibfield  {title} {\bibinfo {title} {Controlling the
  scattering length of ultracold dipolar molecules},\ }\href@noop {} {\bibfield
   {journal} {\bibinfo  {journal} {Phys. Rev. Lett.}\ }\textbf {\bibinfo
  {volume} {121}},\ \bibinfo {pages} {163402} (\bibinfo {year}
  {2018})}\BibitemShut {NoStop}%
\bibitem [{\citenamefont {Anderegg}\ \emph {et~al.}(2021)\citenamefont
  {Anderegg}, \citenamefont {Burchesky}, \citenamefont {Yu}, \citenamefont
  {Karman}, \citenamefont {Chae}, \citenamefont {Ni}, \citenamefont
  {Ketterle},\ and\ \citenamefont {Doyle}}]{Anderegg2021}%
  \BibitemOpen
  \bibfield  {author} {\bibinfo {author} {\bibfnamefont {L.}~\bibnamefont
  {Anderegg}}, \bibinfo {author} {\bibfnamefont {S.}~\bibnamefont {Burchesky}},
  \bibinfo {author} {\bibfnamefont {S.~S.}\ \bibnamefont {Yu}}, \bibinfo
  {author} {\bibfnamefont {T.}~\bibnamefont {Karman}}, \bibinfo {author}
  {\bibfnamefont {E.}~\bibnamefont {Chae}}, \bibinfo {author} {\bibfnamefont
  {K.-K.}\ \bibnamefont {Ni}}, \bibinfo {author} {\bibfnamefont
  {W.}~\bibnamefont {Ketterle}},\ and\ \bibinfo {author} {\bibfnamefont
  {J.~M.}\ \bibnamefont {Doyle}},\ }\bibfield  {title} {\bibinfo {title}
  {Observation of microwave shielding of ultracold molecules},\ }\href@noop {}
  {\bibfield  {journal} {\bibinfo  {journal} {Science}\ }\textbf {\bibinfo
  {volume} {373}},\ \bibinfo {pages} {779} (\bibinfo {year}
  {2021})}\BibitemShut {NoStop}%
\bibitem [{\citenamefont {Schindewolf}\ \emph {et~al.}(2022)\citenamefont
  {Schindewolf}, \citenamefont {Bause}, \citenamefont {Chen}, \citenamefont
  {Duda}, \citenamefont {Karman}, \citenamefont {Bloch},\ and\ \citenamefont
  {Luo}}]{Schindewolf2022}%
  \BibitemOpen
  \bibfield  {author} {\bibinfo {author} {\bibfnamefont {A.}~\bibnamefont
  {Schindewolf}}, \bibinfo {author} {\bibfnamefont {R.}~\bibnamefont {Bause}},
  \bibinfo {author} {\bibfnamefont {X.-Y.}\ \bibnamefont {Chen}}, \bibinfo
  {author} {\bibfnamefont {M.}~\bibnamefont {Duda}}, \bibinfo {author}
  {\bibfnamefont {T.}~\bibnamefont {Karman}}, \bibinfo {author} {\bibfnamefont
  {I.}~\bibnamefont {Bloch}},\ and\ \bibinfo {author} {\bibfnamefont {X.-Y.}\
  \bibnamefont {Luo}},\ }\bibfield  {title} {\bibinfo {title} {Evaporation of
  microwave-shielded polar molecules to quantum degeneracy},\ }\href@noop {}
  {\bibfield  {journal} {\bibinfo  {journal} {Nature}\ }\textbf {\bibinfo
  {volume} {607}},\ \bibinfo {pages} {677} (\bibinfo {year}
  {2022})}\BibitemShut {NoStop}%
\bibitem [{\citenamefont {Chen}\ \emph {et~al.}(2023)\citenamefont {Chen},
  \citenamefont {Schindewolf}, \citenamefont {Eppelt}, \citenamefont {Bause},
  \citenamefont {Duda}, \citenamefont {Biswas}, \citenamefont {Karman},
  \citenamefont {Hilker}, \citenamefont {Bloch},\ and\ \citenamefont
  {Luo}}]{Chen2023}%
  \BibitemOpen
  \bibfield  {author} {\bibinfo {author} {\bibfnamefont {X.-Y.}\ \bibnamefont
  {Chen}}, \bibinfo {author} {\bibfnamefont {A.}~\bibnamefont {Schindewolf}},
  \bibinfo {author} {\bibfnamefont {S.}~\bibnamefont {Eppelt}}, \bibinfo
  {author} {\bibfnamefont {R.}~\bibnamefont {Bause}}, \bibinfo {author}
  {\bibfnamefont {M.}~\bibnamefont {Duda}}, \bibinfo {author} {\bibfnamefont
  {S.}~\bibnamefont {Biswas}}, \bibinfo {author} {\bibfnamefont
  {T.}~\bibnamefont {Karman}}, \bibinfo {author} {\bibfnamefont
  {T.}~\bibnamefont {Hilker}}, \bibinfo {author} {\bibfnamefont
  {I.}~\bibnamefont {Bloch}},\ and\ \bibinfo {author} {\bibfnamefont {X.-Y.}\
  \bibnamefont {Luo}},\ }\bibfield  {title} {\bibinfo {title} {Field-linked
  resonances of polar molecules},\ }\href@noop {} {\bibfield  {journal}
  {\bibinfo  {journal} {Nature}\ }\textbf {\bibinfo {volume} {614}},\ \bibinfo
  {pages} {59} (\bibinfo {year} {2023})}\BibitemShut {NoStop}%
\bibitem [{\citenamefont {Xie}\ \emph {et~al.}(2020)\citenamefont {Xie},
  \citenamefont {Vexiau}, \citenamefont {Orban}, \citenamefont {Dulieu},\ and\
  \citenamefont {Bouloufa-Maafa}}]{Xie2020}%
  \BibitemOpen
  \bibfield  {author} {\bibinfo {author} {\bibfnamefont {M.~L.~T.}\
  \bibnamefont {Xie}}, \bibinfo {author} {\bibfnamefont {R.}~\bibnamefont
  {Vexiau}}, \bibinfo {author} {\bibfnamefont {A.}~\bibnamefont {Orban}},
  \bibinfo {author} {\bibfnamefont {O.}~\bibnamefont {Dulieu}},\ and\ \bibinfo
  {author} {\bibfnamefont {N.}~\bibnamefont {Bouloufa-Maafa}},\ }\bibfield
  {title} {\bibinfo {title} {Optical shielding of destructive chemical
  reactions between ultracold ground-state {NaRb} molecules},\ }\href@noop {}
  {\bibfield  {journal} {\bibinfo  {journal} {Phys. Rev. Lett.}\ }\textbf
  {\bibinfo {volume} {125}},\ \bibinfo {pages} {153202} (\bibinfo {year}
  {2020})}\BibitemShut {NoStop}%
\bibitem [{\citenamefont {Yang}\ \emph {et~al.}(2019)\citenamefont {Yang},
  \citenamefont {Zhang}, \citenamefont {Liu}, \citenamefont {Y.-X.Liu},
  \citenamefont {Nan}, \citenamefont {Zhao},\ and\ \citenamefont
  {Pan}}]{Yang2019}%
  \BibitemOpen
  \bibfield  {author} {\bibinfo {author} {\bibfnamefont {H.}~\bibnamefont
  {Yang}}, \bibinfo {author} {\bibfnamefont {D.-C.}\ \bibnamefont {Zhang}},
  \bibinfo {author} {\bibfnamefont {L.}~\bibnamefont {Liu}}, \bibinfo {author}
  {\bibnamefont {Y.-X.Liu}}, \bibinfo {author} {\bibfnamefont {J.}~\bibnamefont
  {Nan}}, \bibinfo {author} {\bibfnamefont {B.}~\bibnamefont {Zhao}},\ and\
  \bibinfo {author} {\bibfnamefont {J.-W.}\ \bibnamefont {Pan}},\ }\bibfield
  {title} {\bibinfo {title} {Observation of magnetically tunable {F}eshbach
  resonances in ultracold {$^{23}$Na$^{40}$K + $^{40}$K} collisions},\
  }\href@noop {} {\bibfield  {journal} {\bibinfo  {journal} {Science}\ }\textbf
  {\bibinfo {volume} {363}},\ \bibinfo {pages} {261} (\bibinfo {year}
  {2019})}\BibitemShut {NoStop}%
\bibitem [{\citenamefont {Tscherbul}\ and\ \citenamefont
  {Klos}(2020)}]{Tscherbul2020}%
  \BibitemOpen
  \bibfield  {author} {\bibinfo {author} {\bibfnamefont {T.~V.}\ \bibnamefont
  {Tscherbul}}\ and\ \bibinfo {author} {\bibfnamefont {J.}~\bibnamefont
  {Klos}},\ }\bibfield  {title} {\bibinfo {title} {Magnetic tuning of ultracold
  barrierless chemical reactions},\ }\href@noop {} {\bibfield  {journal}
  {\bibinfo  {journal} {Phys. Rev. Res}\ }\textbf {\bibinfo {volume} {2}},\
  \bibinfo {pages} {013117} (\bibinfo {year} {2020})}\BibitemShut {NoStop}%
\bibitem [{\citenamefont {Devolder}\ \emph {et~al.}(2021)\citenamefont
  {Devolder}, \citenamefont {Brumer},\ and\ \citenamefont
  {Tscherbul}}]{Devolder2021}%
  \BibitemOpen
  \bibfield  {author} {\bibinfo {author} {\bibfnamefont {A.}~\bibnamefont
  {Devolder}}, \bibinfo {author} {\bibfnamefont {P.}~\bibnamefont {Brumer}},\
  and\ \bibinfo {author} {\bibfnamefont {T.~V.}\ \bibnamefont {Tscherbul}},\
  }\bibfield  {title} {\bibinfo {title} {Complete quantum coherent control of
  ultracold molecular collisions},\ }\href@noop {} {\bibfield  {journal}
  {\bibinfo  {journal} {Phys. Rev. Lett.}\ }\textbf {\bibinfo {volume} {126}},\
  \bibinfo {pages} {153403} (\bibinfo {year} {2021})}\BibitemShut {NoStop}%
\bibitem [{\citenamefont {Shapiro}\ and\ \citenamefont
  {Brumer}(1996)}]{Shapiro1996}%
  \BibitemOpen
  \bibfield  {author} {\bibinfo {author} {\bibfnamefont {M.}~\bibnamefont
  {Shapiro}}\ and\ \bibinfo {author} {\bibfnamefont {P.}~\bibnamefont
  {Brumer}},\ }\bibfield  {title} {\bibinfo {title} {Coherent control of
  collisional events: Bimolecular reactive scattering},\ }\href@noop {}
  {\bibfield  {journal} {\bibinfo  {journal} {Phys. Rev. Lett.}\ }\textbf
  {\bibinfo {volume} {77}},\ \bibinfo {pages} {2574} (\bibinfo {year}
  {1996})}\BibitemShut {NoStop}%
\bibitem [{\citenamefont {Gong}\ \emph {et~al.}(2003)\citenamefont {Gong},
  \citenamefont {Shapiro},\ and\ \citenamefont {Brumer}}]{Gong2003}%
  \BibitemOpen
  \bibfield  {author} {\bibinfo {author} {\bibfnamefont {J.~B.}\ \bibnamefont
  {Gong}}, \bibinfo {author} {\bibfnamefont {M.}~\bibnamefont {Shapiro}},\ and\
  \bibinfo {author} {\bibfnamefont {P.}~\bibnamefont {Brumer}},\ }\bibfield
  {title} {\bibinfo {title} {Entanglement-assisted coherent control in
  nonreactive diatom-diatom scattering},\ }\href@noop {} {\bibfield  {journal}
  {\bibinfo  {journal} {J. Chem. Phys.}\ }\textbf {\bibinfo {volume} {118}},\
  \bibinfo {pages} {2626} (\bibinfo {year} {2003})}\BibitemShut {NoStop}%
\bibitem [{\citenamefont {Sikorsky}\ \emph {et~al.}(2018)\citenamefont
  {Sikorsky}, \citenamefont {Morita}, \citenamefont {Meir}, \citenamefont
  {Buchachenko}, \citenamefont {Ben-Shlomi}, \citenamefont {Akerman},
  \citenamefont {Narevicius}, \citenamefont {Tscherbul},\ and\ \citenamefont
  {Ozeri}}]{Sikorsky2018}%
  \BibitemOpen
  \bibfield  {author} {\bibinfo {author} {\bibfnamefont {T.}~\bibnamefont
  {Sikorsky}}, \bibinfo {author} {\bibfnamefont {M.}~\bibnamefont {Morita}},
  \bibinfo {author} {\bibfnamefont {Z.}~\bibnamefont {Meir}}, \bibinfo {author}
  {\bibfnamefont {A.~A.}\ \bibnamefont {Buchachenko}}, \bibinfo {author}
  {\bibfnamefont {R.}~\bibnamefont {Ben-Shlomi}}, \bibinfo {author}
  {\bibfnamefont {N.}~\bibnamefont {Akerman}}, \bibinfo {author} {\bibfnamefont
  {E.}~\bibnamefont {Narevicius}}, \bibinfo {author} {\bibfnamefont {T.~V.}\
  \bibnamefont {Tscherbul}},\ and\ \bibinfo {author} {\bibfnamefont
  {R.}~\bibnamefont {Ozeri}},\ }\bibfield  {title} {\bibinfo {title} {Phase
  locking between different partial waves in atom-ion spin-exchange
  collisions},\ }\href@noop {} {\bibfield  {journal} {\bibinfo  {journal}
  {Phys. Rev. Lett.}\ }\textbf {\bibinfo {volume} {121}},\ \bibinfo {pages}
  {173402} (\bibinfo {year} {2018})}\BibitemShut {NoStop}%
\bibitem [{\citenamefont {Kartz}\ \emph {et~al.}(2022)\citenamefont {Kartz},
  \citenamefont {Pinkas}, \citenamefont {Akerman},\ and\ \citenamefont
  {Ozeri}}]{Kartz2022}%
  \BibitemOpen
  \bibfield  {author} {\bibinfo {author} {\bibfnamefont {O.}~\bibnamefont
  {Kartz}}, \bibinfo {author} {\bibfnamefont {M.}~\bibnamefont {Pinkas}},
  \bibinfo {author} {\bibfnamefont {N.}~\bibnamefont {Akerman}},\ and\ \bibinfo
  {author} {\bibfnamefont {R.}~\bibnamefont {Ozeri}},\ }\bibfield  {title}
  {\bibinfo {title} {Quantum logic detection of collisions between single
  atom–ion pairs},\ }\href@noop {} {\bibfield  {journal} {\bibinfo  {journal}
  {Nat. Phys.}\ }\textbf {\bibinfo {volume} {18}},\ \bibinfo {pages} {533}
  (\bibinfo {year} {2022})}\BibitemShut {NoStop}%
\bibitem [{\citenamefont {Tomza}\ \emph {et~al.}(2019)\citenamefont {Tomza},
  \citenamefont {Jachymski}, \citenamefont {Gerritsma}, \citenamefont
  {Negretti}, \citenamefont {Calarco}, \citenamefont {Idziaszek},\ and\
  \citenamefont {Julienne}}]{Tomza2019}%
  \BibitemOpen
  \bibfield  {author} {\bibinfo {author} {\bibfnamefont {M.}~\bibnamefont
  {Tomza}}, \bibinfo {author} {\bibfnamefont {K.}~\bibnamefont {Jachymski}},
  \bibinfo {author} {\bibfnamefont {R.}~\bibnamefont {Gerritsma}}, \bibinfo
  {author} {\bibfnamefont {A.}~\bibnamefont {Negretti}}, \bibinfo {author}
  {\bibfnamefont {T.}~\bibnamefont {Calarco}}, \bibinfo {author} {\bibfnamefont
  {Z.}~\bibnamefont {Idziaszek}},\ and\ \bibinfo {author} {\bibfnamefont
  {P.~S.}\ \bibnamefont {Julienne}},\ }\bibfield  {title} {\bibinfo {title}
  {Cold hybrid ion-atom systems},\ }\href@noop {} {\bibfield  {journal}
  {\bibinfo  {journal} {Rev. Mod. Phys.}\ }\textbf {\bibinfo {volume} {91}},\
  \bibinfo {pages} {035001} (\bibinfo {year} {2019})}\BibitemShut {NoStop}%
\bibitem [{\citenamefont {Coté}\ and\ \citenamefont
  {Simbotin}(2018)}]{Cote2018}%
  \BibitemOpen
  \bibfield  {author} {\bibinfo {author} {\bibfnamefont {R.}~\bibnamefont
  {Coté}}\ and\ \bibinfo {author} {\bibfnamefont {I.}~\bibnamefont
  {Simbotin}},\ }\bibfield  {title} {\bibinfo {title} {Signature of the s-wave
  regime high above ultralow temperatures},\ }\href@noop {} {\bibfield
  {journal} {\bibinfo  {journal} {Phys. Rev. Lett.}\ }\textbf {\bibinfo
  {volume} {121}},\ \bibinfo {pages} {173401} (\bibinfo {year}
  {2018})}\BibitemShut {NoStop}%
\bibitem [{\citenamefont {Gao}(2001)}]{Gao2001}%
  \BibitemOpen
  \bibfield  {author} {\bibinfo {author} {\bibfnamefont {B.}~\bibnamefont
  {Gao}},\ }\bibfield  {title} {\bibinfo {title} {Angular-momentum-insensitive
  quantum-defect theory for diatomic systems},\ }\href@noop {} {\bibfield
  {journal} {\bibinfo  {journal} {Phys. Rev. A}\ }\textbf {\bibinfo {volume}
  {64}},\ \bibinfo {pages} {010701(R)} (\bibinfo {year} {2001})}\BibitemShut
  {NoStop}%
\bibitem [{\citenamefont {Omiste}\ \emph {et~al.}(2018)\citenamefont {Omiste},
  \citenamefont {Flo\ss},\ and\ \citenamefont {Brumer}}]{Omiste2018}%
  \BibitemOpen
  \bibfield  {author} {\bibinfo {author} {\bibfnamefont {J.~J.}\ \bibnamefont
  {Omiste}}, \bibinfo {author} {\bibfnamefont {J.}~\bibnamefont {Flo\ss}},\
  and\ \bibinfo {author} {\bibfnamefont {P.}~\bibnamefont {Brumer}},\
  }\bibfield  {title} {\bibinfo {title} {Coherent control of penning and
  associative ionization: Insights from symmetries},\ }\href@noop {} {\bibfield
   {journal} {\bibinfo  {journal} {Phys. Rev. Lett.}\ }\textbf {\bibinfo
  {volume} {121}},\ \bibinfo {pages} {163405} (\bibinfo {year}
  {2018})}\BibitemShut {NoStop}%
\bibitem [{\citenamefont {Devolder}\ \emph {et~al.}(2022)\citenamefont
  {Devolder}, \citenamefont {Tscherbul},\ and\ \citenamefont
  {Brumer}}]{Devolder2022}%
  \BibitemOpen
  \bibfield  {author} {\bibinfo {author} {\bibfnamefont {A.}~\bibnamefont
  {Devolder}}, \bibinfo {author} {\bibfnamefont {T.~V.}\ \bibnamefont
  {Tscherbul}},\ and\ \bibinfo {author} {\bibfnamefont {P.}~\bibnamefont
  {Brumer}},\ }\bibfield  {title} {\bibinfo {title} {Coherent multichannel
  optical theorem: Quantum control of the total scattering cross section},\
  }\href@noop {} {\bibfield  {journal} {\bibinfo  {journal} {Phys. Rev. A}\
  }\textbf {\bibinfo {volume} {105}},\ \bibinfo {pages} {052808} (\bibinfo
  {year} {2022})}\BibitemShut {NoStop}%
\bibitem [{\citenamefont {Devolder}\ \emph {et~al.}(2023)\citenamefont
  {Devolder}, \citenamefont {Tscherbul},\ and\ \citenamefont
  {Brumer}}]{Devolder2023}%
  \BibitemOpen
  \bibfield  {author} {\bibinfo {author} {\bibfnamefont {A.}~\bibnamefont
  {Devolder}}, \bibinfo {author} {\bibfnamefont {T.~V.}\ \bibnamefont
  {Tscherbul}},\ and\ \bibinfo {author} {\bibfnamefont {P.}~\bibnamefont
  {Brumer}},\ }\bibfield  {title} {\bibinfo {title} {Coherent control of
  ultracold molecular collisions: The role of resonances},\ }\href@noop {}
  {\bibfield  {journal} {\bibinfo  {journal} {J. Phys. Chem. Lett.}\ }\textbf
  {\bibinfo {volume} {14}},\ \bibinfo {pages} {2171} (\bibinfo {year}
  {2023})}\BibitemShut {NoStop}%
\end{thebibliography}%
\end{document}